\documentclass[%
 reprint,
superscriptaddress,
 amsmath,amssymb,
 aps,
 longbibliography,
]{revtex4-1}

\usepackage{graphicx}
\usepackage{dcolumn}
\usepackage{bm}
\usepackage{amsmath,amssymb,amstext,amsthm}
\usepackage{braket}
\usepackage{bbold}
\usepackage{bbm}
\usepackage{xcolor}
\usepackage{ gensymb }
\usepackage{soul}
\usepackage{enumitem}

\begin{document}

\preprint{APS/123-QED}

\title{Generation and detection of spin-orbit coupled neutron beams}

\author{D. Sarenac}
\email{dsarenac@uwaterloo.ca}
\affiliation{Institute for Quantum Computing, University of Waterloo,  Waterloo, ON, Canada, N2L3G1}
\author{C. Kapahi}
\affiliation{Institute for Quantum Computing, University of Waterloo,  Waterloo, ON, Canada, N2L3G1}
\affiliation{Department of Physics, University of Waterloo, Waterloo, ON, Canada, N2L3G1}
\author{W. C. Chen}
\affiliation{National Institute of Standards and Technology, Gaithersburg, Maryland 20899, USA}
\affiliation{University of Maryland, College Park, Maryland 20742, USA}
\author{Charles W. Clark}
\affiliation{Joint Quantum Institute, National Institute of Standards and Technology and University of Maryland, College Park, Maryland 20742, USA}
\author{D. G. Cory}
\affiliation{Institute for Quantum Computing, University of Waterloo,  Waterloo, ON, Canada, N2L3G1} 
\affiliation{Department of Chemistry, University of Waterloo, Waterloo, ON, Canada, N2L3G1}
\affiliation{Perimeter Institute for Theoretical Physics, Waterloo, ON, Canada, N2L2Y5}
\affiliation{Canadian Institute for Advanced Research, Toronto, Ontario, Canada, M5G 1Z8}
\author{M. G. Huber}
\affiliation{National Institute of Standards and Technology, Gaithersburg, Maryland 20899, USA}
\author{I. Taminiau}
\affiliation{Institute for Quantum Computing, University of Waterloo,  Waterloo, ON, Canada, N2L3G1}
\author{K. Zhernenkov}
\affiliation{Institute for Quantum Computing, University of Waterloo,  Waterloo, ON, Canada, N2L3G1}
\affiliation{J\"ulich Centre for Neutron Science at Heinz Maier-Leibnitz Zentrum, Forschungszentrum J\"ulich GmbH, 85748 Garching, Germany}
\author{D. A. Pushin}
\email{dmitry.pushin@uwaterloo.ca}
\affiliation{Institute for Quantum Computing, University of Waterloo,  Waterloo, ON, Canada, N2L3G1}
\affiliation{Department of Physics, University of Waterloo, Waterloo, ON, Canada, N2L3G1}

\date{\today}

\begin{abstract}

Spin-orbit coupling of light has come to the fore in nano-optics and plasmonics, and is a key ingredient of topological photonics and chiral quantum optics. We demonstrate a basic tool for incorporating analogous effects into neutron optics: the generation and detection of neutron beams with coupled spin and orbital angular momentum. $^3$He neutron spin-filters are used in conjunction with specifically oriented triangular coils to prepare neutron beams with lattices of spin-orbit correlations, as demonstrated by their spin-dependant intensity profiles. These correlations can be tailored to particular applications, such as neutron studies of topological materials. 

\end{abstract}

\maketitle

\section{Introduction}

Studies of optical OAM have blossomed since the early 1990's and are now encompassed in a larger framework of structured waves of light and matter \cite{rubinsztein2016roadmap,BarnettBabikerPadgett}. OAM 
has been induced in beams of light \cite{LesAllen1992}, electrons \cite{bliokh2007semiclassical,uchida2010generation,mcmorran2011electron}, and neutrons \cite{Dima2015}. 
Photonic OAM has demonstrated usefulness in edge-detection microscopy, quantum information processing protocols, encoding and multiplexing of communications, and optical manipulation of matter. \cite{mair2001entanglement,wang2012terabit,Andersen2006,he1995direct,friese1996optical,brullot2016resolving,Simpson:97,Zhang:19}. Electron OAM beams have found applications in the characterization of nanoscale magnetic fields in materials
~\cite{grillo2017observation} and exploration of magnetic monopoles~\cite{beche2014magnetic}. Neutron OAM has shown promise in the detection of buried interfaces
~\cite{holography}. Furthermore, it has been theorized that neutron OAM can modify Schwinger scattering of neutrons on nuclei~\cite{SchwingerOam}, and might also enable studies of neutron's internal structure~\cite{larocque2018twisting}.

A related set of techniques have been developed for preparing and characterizing beams in which the spin and OAM are correlated. In the case of photons, these ``spin-orbit'' beams possess correlations between polarization and the OAM \cite{maurer2007tailoring,qplate}, whereas for electrons and neutrons the correlations are between the spin and the OAM \cite{karimi2012spin,spinorbit}. Photonic spin-orbit beams have been demonstrated and they have enriched the application range of OAM beams by increasing the number of accessible degrees of freedom  \cite{marrucci2011spin,milione20154,schmiegelow2016transfer,vallone2014freespace}. Although analogous preparation methods have been proposed for electrons~\cite{karimi2012spin,karimi2014generation} they have yet to be implemented in the laboratory.


In this paper we demonstrate not only the first preparation and characterization of spin-orbit beams using neutrons, but also neutron beams with a lattice of spin-orbit correlations. Our technique, which we previously demonstrated with light \cite{sarenac2018generation}, involves the preparation of a spin-orbit textured ``lattice of vortices'' wavefront. A variety of textures can be generated by this method \cite{methods}, including skyrmion-like geometries analogous to those recently observed in evanescent electromagnetic fields \cite{2018Sci...361..993T}. We expect the techniques shown here to pave the way for neutron OAM and spin-orbit applications in material characterization and fundamental physics.



\section{Methods}

\begin{figure*}
\centering\includegraphics[width=\linewidth]{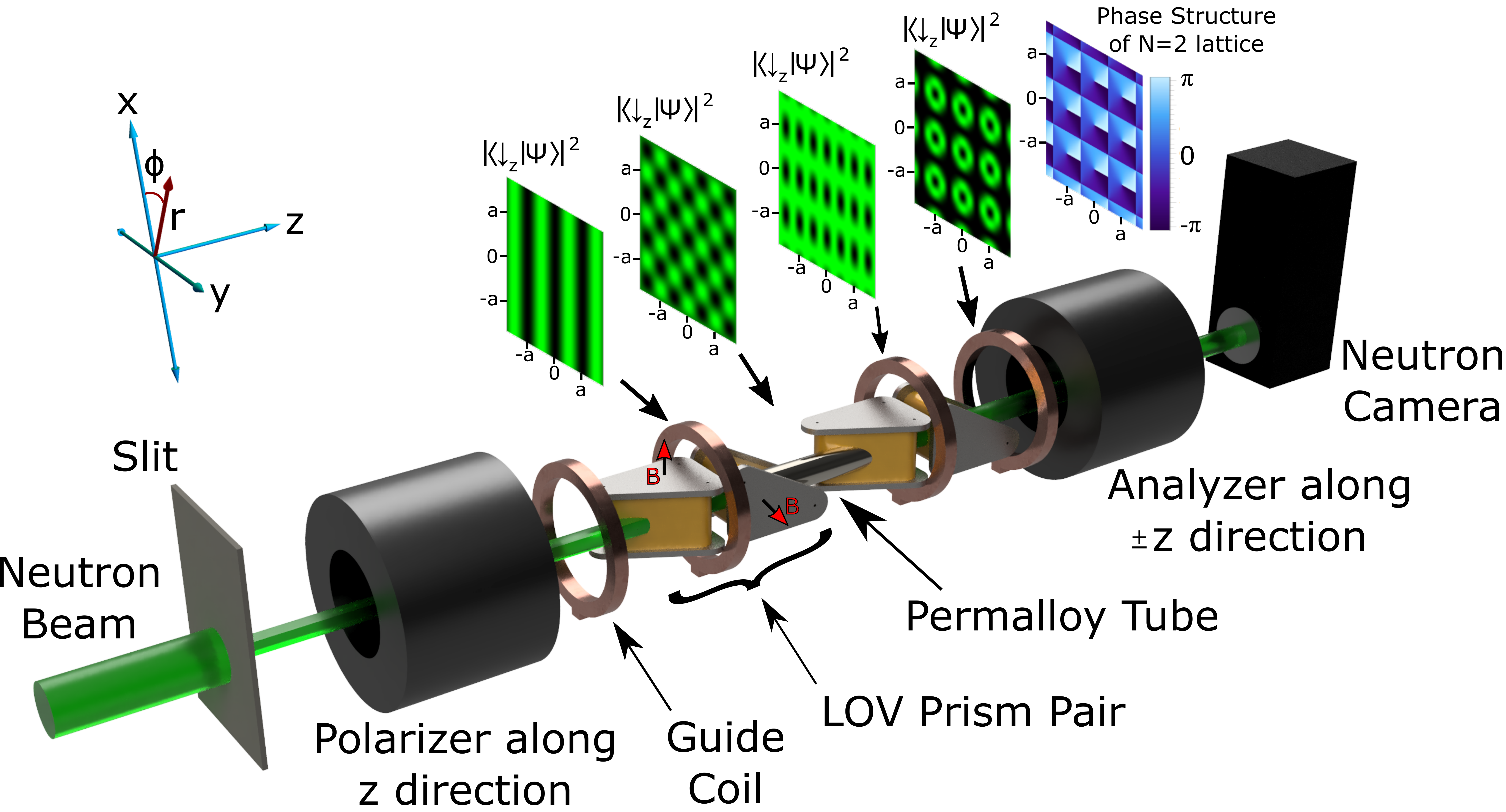}
\caption{Schematic diagram of the experimental setup which consists of a slit, a $^3$He spin polarizer and analyzer, guide field coils, two pairs of specifically oriented triangular coils that act as LOV prism pairs, a permalloy tube, and a neutron camera. The triangular coils induce perpendicular magnetic phase gradients onto the neutron wavefunction. LOV prism pairs prepare beams with a lattice of spin-orbit correlations where in each lattice cell the phase between the two spin states varies azimuthally. The simulated spin dependant intensity profile after each triangular coil is shown in green, and the profile of the phase difference between the two spin states, given by $\left(\text{arg}[\braket{\downarrow_z|\Psi_\mathrm{LOV}^{N=2}}]-\text{arg}[\braket{\uparrow_z|\Psi_\mathrm{LOV}^{N=2}}]\right)$,  is shown in blue.}
 \label{fig:setup}
\end{figure*}

The experiment was carried out the on the Polarized $^3$He And Detector Experiment Station (PHADES)~\cite{phades} at the National Institute for Standards and Technology Center for Neutron Research (NCNR). A monochromatic 
beam of neutrons with wavelength $\lambda=0.41$ nm  was directed into the setup, as shown in Fig.~\ref{fig:setup}. The beam divergence was $\sim1\degree$ in both $x$ and $y$ directions. The setup is composed of a slit, two $^3$He neutron spin filters, guide field coils, two pairs of specifically orientated triangular coils, a permalloy tube, and a neutron camera. The neutron camera has a 25 mm diameter active area and a spatial resolution of 100 $\mu$m~\cite{dietze1996intensified}.

A $1\times1$ mm$^2$ slit was placed at the start of the setup. The slit sets the lower limit on the transverse coherence length at the first triangular coil to:
\begin{align}
\sigma=\frac{\lambda L_1}{s}\sim0.4~\mu \text{m} 
\label{eq:coherence}
\end{align}

\noindent where $L_1=0.965$ m is the distance from the slit to the first triangular coil and $s$ is the slit width. Although it might be desirable for some specific applications to reduce the slit width to the point that the transverse coherence length extends over the beam diameter, it is not practical in this experiment as the neutron peak count rate was $\sim15$~$s^{-1}$ at the camera. 

Two $^3$He cells were used as the spin polarizer and the spin analyzer due to their spatially homogeneous neutron polarization~\cite{chen20143he}. The cells were polarized in an off-line lab using spin-exchange optical pumping~\cite{chen2014limits}, and they were changed three times during the experiment.  Their initial $^3$He polarization at the beamline was measured to be between $73~\%$ and $82~\%$, while their relaxation time was measured to be between $365$ h and $516$ h. The polarization of the neutron beam would reduce from $\sim~94~\%$ to $\sim~90~\%$ during a 2-3 day time period. The spin filter direction could be aligned with the $\pm z-$axis at the beamline using the adiabatic fast passage nuclear magnetic resonance method~\cite{abragam1961principles}.

\begin{figure*}
\centering\includegraphics[width=\linewidth]{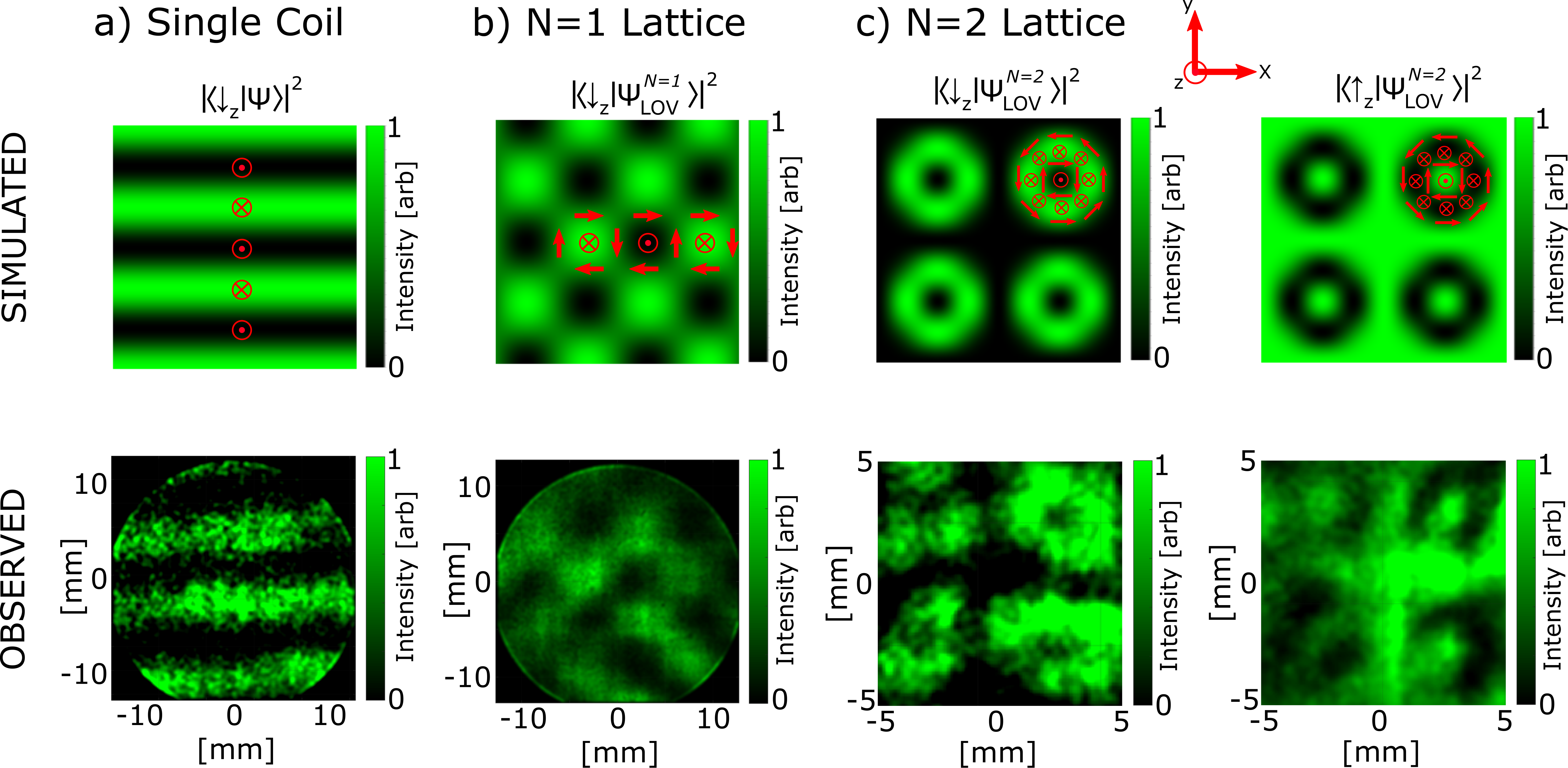}
\caption{The simulated and observed spin-dependant intensity profiles. A Gaussian filter as well as an intensity gradient was added to each observed image, to highlight the features of interest. The currents on the (1$^{st}$, 2$^{nd}$, 3$^{rd}$, 4$^{th}$) triangular coil were set to (0, 0, 0, 2.5 A) for a), (2.5, 2.5, 0, 0 A) for b), and (5, 5, 5, 4 A) for c). The spatially varying spin direction (before the spin filtering) is overlaid on the simulated intensity profiles via the red arrows. The N=1 lattice exhibits a vortex anti-vortex structure and its spin dependant intensity profile resembles a checkerboard pattern. The N=2 lattice appears as a lattice of doughnut/ring shapes. Good qualitative agreement is shown between the simulated and observed intensity profiles. }
 \label{fig:N1}
\end{figure*}

To prepare neutron beams with lattices of spin-orbit correlations we used specifically orientated triangular coils that acted as ``Lattice of Optical Vortices'' (LOV) prism pairs \cite{sarenac2018generation,methods}. They were arranged to induce magnetic phase gradients perpendicular to each other as well as to the incoming neutron beam. Therefore the coil arrangement differs from the Wollaston arrangement where two triangular coils with anti-parallel fields are placed with their inclined sides facing each other~\cite{li2017high,bouwman2008real}. The triangular coils have side lengths of 8.5~cm, 12~cm, and 14.7~cm with an overall height of 7.3~cm. At an applied 10~A current their inner magnetic field was $\sim 0.014$~T which provided a magnetic phase gradient of $\sim1.5\pi$~rad/mm. The triangular coils were run between 2.5~A and 10~A throughout the experiment, and in every configuration the current in each coil was optimized to compensate for beam divergence.

As shown in Fig.~\ref{fig:setup} there is a spatially dependant path difference between the 2$^\text{nd}$ and the 3$^\text{rd}$ triangular coils due to their inclined sides. Therefore it is necessary to minimize the magnetic field in this region to avoid an unwanted phase gradient across the beam. In our setup this was accomplished via a permalloy tube. The tube was built from 15 layers of a $500$ $\mu$m thick nickel-iron soft ferromagnetic sheets. The sheets were wrapped around a thin-walled aluminium pipe with an inner diameter of 3.18 cm and whose ends were cut to match the angled prism faces. Guide coils were placed between other triangular coils to provide a uniform magnetic field along the spin quantization axis.

\section{Results and Discussion}

The first $^3$He polarizer filters the neutrons with spin along the beam propagation axis, thereby setting the neutron wavefunction to

\begin{align}
    \ket{\Psi_\mathrm{in}}=	\ket{\uparrow_z}.
	\label{Eqn:up}
\end{align}

\noindent The triangular coils induce perpendicular phase gradients along the directions that are also perpendicular to the direction of the incoming spin state. Pairs of triangular coils then effectively act as LOV prism pairs, as described in Ref.~\cite{methods}. In this particular case their individual operators are given by:

\begin{align}
	\hat{U}_{y}=e^{-i\frac{\pi }{a}y\hat{\sigma}_x} \qquad 
	\hat{U}_{x}=e^{-i\frac{\pi }{a}x\hat{\sigma}_y}
	\label{gradientoperators}
\end{align}

\noindent where $\hat{\sigma}_y$ and $\hat{\sigma}_x$ are the Pauli spin operators, and $a$ is the spatial spin oscillation period. For the case of no beam divergence:

\begin{align}
	a=\frac{2\pi v_z}{ \gamma_\mathrm{n} |B|  \tan(\theta)}
	\label{Eqn:const}
\end{align}

\noindent where $|B|$ is the magnetic field inside the triangular coils, $v_z$ is the neutron velocity, $\gamma_\mathrm{n}$ is the neutron gyromagnetic ratio~\cite{codata}, and $\theta$ is the incline angle of the triangle coils. For example, for a field of $|B|\sim0.005$ T inside the triangular coils the corresponding period of a non-diverging beam would be $a\sim3.8$ mm.

A pair of specifically orientated triangular coils, or a LOV prism pair, approximates the action of a quadrupole magnetic field~\cite{sarenac2018generation}. The state induced by a quadrople acting on $\ket{\Psi_\mathrm{in}}=	\ket{\uparrow_z}$ has the following form~\cite{spinorbit}:

\begin{align}
	\ket{\Psi_Q}
    	\sim
        	\left[
            	\cos\left(\frac{\pi r}{a}\right) \ket{\uparrow_z}
                +ie^{-i\phi}\sin\left(\frac{\pi r}{a}\right)\ket{\downarrow_z}
            \right],
	\label{Eqn:psiq}
\end{align}
\noindent where $(r,\phi)$ are the cylindrical coordinates. It follows from Eq.~\ref{Eqn:psiq} that two spin states possess a differing spatial amplitude profile and that there is an azimuthal phase difference between the two spin states which indicates the OAM difference between the two spin states of $\Delta\ell=\ell_\uparrow-\ell_\downarrow=1$.

In addition to approximating the quadrupole operator, LOV prism pairs possess a periodic structure which induces a 2D lattice structure in the output state~\cite{sarenac2018generation}. The state after N sets of LOV prism pairs is given by:

\begin{align}
	\ket{\Psi_\mathrm{LOV}^N}
    	=(\hat{U}_x \hat{U}_y)^N\ket{\Psi_\mathrm{in}}.
	\label{Eqn:psiLOV}
\end{align}

After passing through one of the triangular coils the spin polarization of the beam oscillates along the direction of the coil incline. Therefore the intensity profile post-selected on one spin state exhibits linear fringes with period $a$, as shown in Fig.~\ref{fig:N1}a. After passing through a pair of perpendicular triangular coils, or a LOV prism pair, an N=1 lattice of spin-orbit correlations is prepared. The intensity profile post-selected on $\ket{\downarrow_z}$ is shown in Fig.~\ref{fig:N1}b, and it resembles a checkerboard pattern. The spin direction before the post-selection, is overlaid on the intensity profile via the red arrows and it elucidates why the N=1 lattice is is composed of a vortex anti-vortex structure. 

\begin{figure}
\centering\includegraphics[width=\linewidth]{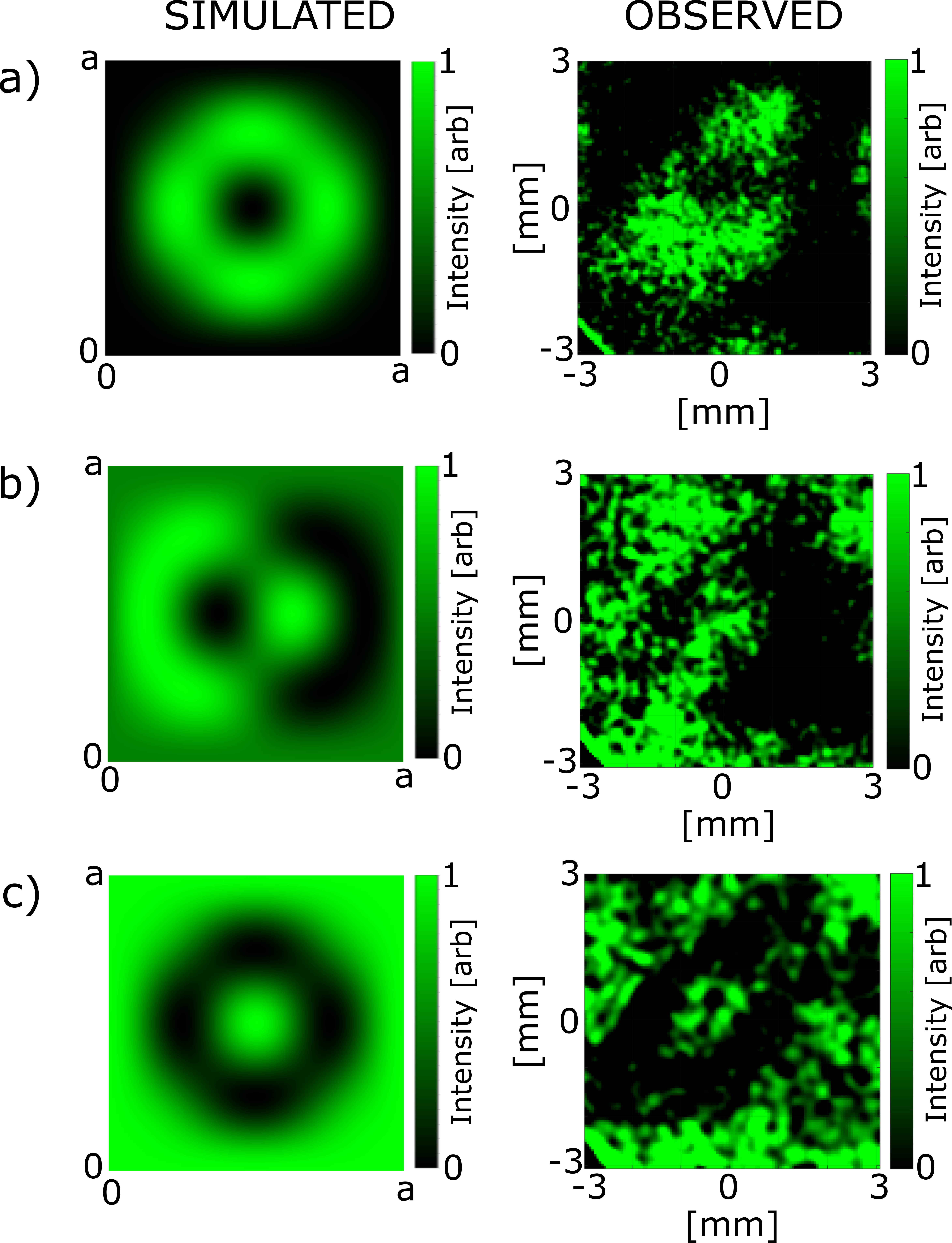}
\caption{The simulated and observed spin dependant intensity profile, after two sets of LOV prism pairs, as the first coil in the setup is translated along the y-direction, see Fig.~\ref{fig:setup}, for a) 0 mm b) 3 mm c) 6 mm. The profiles of a) and c) correspond to the spin-up and spin down intensity profiles of a single cell of the N=2 lattice, as shown in Fig.~\ref{fig:N1}. The profile in b) corresponds to the intensity profile after mixing the two spin states. It can be seen that the intensity varies azimuthally, indicating the phase structure of a single cell in the N=2 lattice, as shown in Fig.~\ref{fig:setup}.  }
 \label{fig:phase}
\end{figure}

Passing a polarized neutron beam through two pairs of LOV prisms pairs prepares a beam with a lattice of spin-orbit correlations as described by Eq.~\ref{Eqn:psiq}. The spin dependant intensity profile has the doughnut/ring structure as shown in Fig.~\ref{fig:N1}c. This is a consequence of the cosine/sine amplitude terms in Eq.~\ref{Eqn:psiq}. The major features can be seen between the simulated and observed profiles in Fig.~\ref{fig:N1}c. Note that the spin analyzer sets the spin filter direction, and the two profiles in Fig.~\ref{fig:N1}c are from two separate setup configurations.

The slight differences between the simulated and observed profiles shown in Fig.~\ref{fig:N1}a and Fig.~\ref{fig:N1}b can be attributed to the interface region between the longitudinal field of the guide coils and the transverse field of the triangular coils. However, when triangular coils 2 and 3 are used to prepare the N=1 lattice the observed profile is significantly more distorted, indicating that the permalloy tube is not sufficiently removing the field between the triangular coils.

The phase difference between the two spin states of the N=2 lattice is shown in Fig.~\ref{fig:setup}. This phase structure can be mapped via the spin-dependant intensity profile after mixing the two spin states. That is, we require to post-select the spin along a direction that is perpendicular to the spin quantization axis, which in our case would be the x and y directions. 

It can be noted that translating one of the triangular coils along its incline direction induces an additional uniform phase shift. This provides a convenient method of obtaining the $\ket{\downarrow_\text{x,y}}$ dependant intensity profiles without changing the $^3$He polarization direction~\cite{methods}. Fig.~\ref{fig:phase} shows the simulated and observed spin dependant intensity profile as the first coil in the setup is translated along the y-direction. Fig.~\ref{fig:phase}a and Fig.~\ref{fig:phase}c correspond to the spin-up and spin down intensity profiles of a single cell of the N=2 lattice, as shown in Fig.~\ref{fig:N1}. Fig.~\ref{fig:phase}b corresponds to the intensity profile after mixing the two spin states. It can be seen that the intensity varies azimuthally, indicating the phase structure of the N=2 lattice as shown in Fig.~\ref{fig:setup}.

\section{Conclusion}

Photon spin-orbit coupling arises naturally in nano-optics, photonics, plasmonics and optical metamaterials~\cite{bliokh2015spin,Stav1101} and is a core construct of chiral quantum optics~\cite{lodahl2017chiral} and topological photonics~\cite{ozawa2019topological}. In this work we explore spin-orbit coupling in the context of freely-propagating beams in which spin and orbital angular momentum (OAM) degrees of freedom are correlated. We have prepared and characterized neutron beams with lattices of spin-orbit correlations in which the OAM of one spin state is different from the OAM of the other spin state. This was achieved via sets of specifically oriented triangular coils which acted as LOV prism pairs. The beams were characterized via their spin dependant intensity profiles. 

The triangular coils induced good quality magnetic phase gradients, as can be observed in Fig.~\ref{fig:N1}a and Fig.~\ref{fig:N1}b. However, in our experiment the permalloy tube did not sufficiently remove the magnetic field between the two sets of triangular coils. This resulted in distortions when all four coils were on simultaneously. For more pronounced results a better mechanism of removing the field is required. 

We expect the described techniques to be the forerunners of neutron OAM applications in material characterization and fundamental physics. Superconducting triangular coils with higher fields may be employed to prepare lattices with smaller periods. The next set of experiments will focus on the preparation of spin-orbit correlations over the coherence length of neutron wave packets and the characterization of these spin-orbit states via the correlations between spin and projected linear momentum.

\section{Acknowledgements}

This work was supported by the Canadian Excellence Research Chairs (CERC) program, the Natural Sciences and Engineering Research Council of Canada (NSERC) Discovery program, Collaborative Research and Training Experience (CREATE) program, the Canada  First  Research  Excellence  Fund  (CFREF), and the National Institute of Standards and Technology (NIST) Quantum Information Program. The work utilized facilities supported in part by the National Science Foundation under Agreement No. DMR-1508249. The authors thank S. Watson and M. T. Hassan for their help in the experimental setup on PHADES.

\bibliography{OAMrefs}

\begin{thebibliography}{44}%
\makeatletter
\providecommand \@ifxundefined [1]{%
 \@ifx{#1\undefined}
}%
\providecommand \@ifnum [1]{%
 \ifnum #1\expandafter \@firstoftwo
 \else \expandafter \@secondoftwo
 \fi
}%
\providecommand \@ifx [1]{%
 \ifx #1\expandafter \@firstoftwo
 \else \expandafter \@secondoftwo
 \fi
}%
\providecommand \natexlab [1]{#1}%
\providecommand \enquote  [1]{``#1''}%
\providecommand \bibnamefont  [1]{#1}%
\providecommand \bibfnamefont [1]{#1}%
\providecommand \citenamefont [1]{#1}%
\providecommand \href@noop [0]{\@secondoftwo}%
\providecommand \href [0]{\begingroup \@sanitize@url \@href}%
\providecommand \@href[1]{\@@startlink{#1}\@@href}%
\providecommand \@@href[1]{\endgroup#1\@@endlink}%
\providecommand \@sanitize@url [0]{\catcode `\\12\catcode `\$12\catcode
  `\&12\catcode `\#12\catcode `\^12\catcode `\_12\catcode `\%12\relax}%
\providecommand \@@startlink[1]{}%
\providecommand \@@endlink[0]{}%
\providecommand \url  [0]{\begingroup\@sanitize@url \@url }%
\providecommand \@url [1]{\endgroup\@href {#1}{\urlprefix }}%
\providecommand \urlprefix  [0]{URL }%
\providecommand \Eprint [0]{\href }%
\providecommand \doibase [0]{http://dx.doi.org/}%
\providecommand \selectlanguage [0]{\@gobble}%
\providecommand \bibinfo  [0]{\@secondoftwo}%
\providecommand \bibfield  [0]{\@secondoftwo}%
\providecommand \translation [1]{[#1]}%
\providecommand \BibitemOpen [0]{}%
\providecommand \bibitemStop [0]{}%
\providecommand \bibitemNoStop [0]{.\EOS\space}%
\providecommand \EOS [0]{\spacefactor3000\relax}%
\providecommand \BibitemShut  [1]{\csname bibitem#1\endcsname}%
\let\auto@bib@innerbib\@empty
\bibitem [{\citenamefont {Rubinsztein-Dunlop}\ \emph
  {et~al.}(2016)\citenamefont {Rubinsztein-Dunlop}, \citenamefont {Forbes},
  \citenamefont {Berry}, \citenamefont {Dennis}, \citenamefont {Andrews},
  \citenamefont {Mansuripur}, \citenamefont {Denz}, \citenamefont {Alpmann},
  \citenamefont {Banzer}, \citenamefont {Bauer} \emph
  {et~al.}}]{rubinsztein2016roadmap}%
  \BibitemOpen
  \bibfield  {author} {\bibinfo {author} {\bibfnamefont {Halina}\ \bibnamefont
  {Rubinsztein-Dunlop}}, \bibinfo {author} {\bibfnamefont {Andrew}\
  \bibnamefont {Forbes}}, \bibinfo {author} {\bibfnamefont {MV}~\bibnamefont
  {Berry}}, \bibinfo {author} {\bibfnamefont {MR}~\bibnamefont {Dennis}},
  \bibinfo {author} {\bibfnamefont {David~L}\ \bibnamefont {Andrews}}, \bibinfo
  {author} {\bibfnamefont {Masud}\ \bibnamefont {Mansuripur}}, \bibinfo
  {author} {\bibfnamefont {Cornelia}\ \bibnamefont {Denz}}, \bibinfo {author}
  {\bibfnamefont {Christina}\ \bibnamefont {Alpmann}}, \bibinfo {author}
  {\bibfnamefont {Peter}\ \bibnamefont {Banzer}}, \bibinfo {author}
  {\bibfnamefont {Thomas}\ \bibnamefont {Bauer}},  \emph {et~al.},\ }\bibfield
  {title} {\enquote {\bibinfo {title} {Roadmap on structured light},}\
  }\href@noop {} {\bibfield  {journal} {\bibinfo  {journal} {Journal of
  Optics}\ }\textbf {\bibinfo {volume} {19}},\ \bibinfo {pages} {013001}
  (\bibinfo {year} {2016})}\BibitemShut {NoStop}%
\bibitem [{\citenamefont {Barnett}\ \emph {et~al.}(2017)\citenamefont
  {Barnett}, \citenamefont {Babiker},\ and\ \citenamefont
  {Padgett}}]{BarnettBabikerPadgett}%
  \BibitemOpen
  \bibfield  {author} {\bibinfo {author} {\bibfnamefont {Stephen~M.}\
  \bibnamefont {Barnett}}, \bibinfo {author} {\bibfnamefont {Mohamed}\
  \bibnamefont {Babiker}}, \ and\ \bibinfo {author} {\bibfnamefont {Miles~J.}\
  \bibnamefont {Padgett}},\ }\bibfield  {title} {\enquote {\bibinfo {title}
  {Optical orbital angular momentum},}\ }\href
  {https://doi.org/10.1098/rsta.2015.0444} {\bibfield  {journal} {\bibinfo
  {journal} {Philosophical Transactions of the Royal Society A}\ }\textbf
  {\bibinfo {volume} {375}} (\bibinfo {year} {2017})}\BibitemShut {NoStop}%
\bibitem [{\citenamefont {Allen}\ \emph {et~al.}(1992)\citenamefont {Allen},
  \citenamefont {Beijersbergen}, \citenamefont {Spreeuw},\ and\ \citenamefont
  {Woerdman}}]{LesAllen1992}%
  \BibitemOpen
  \bibfield  {author} {\bibinfo {author} {\bibfnamefont {L.}~\bibnamefont
  {Allen}}, \bibinfo {author} {\bibfnamefont {M.~W.}\ \bibnamefont
  {Beijersbergen}}, \bibinfo {author} {\bibfnamefont {R.~J.~C.}\ \bibnamefont
  {Spreeuw}}, \ and\ \bibinfo {author} {\bibfnamefont {J.~P.}\ \bibnamefont
  {Woerdman}},\ }\bibfield  {title} {\enquote {\bibinfo {title} {Orbital
  angular momentum of light and the transformation of laguerre-gaussian laser
  modes},}\ }\href {\doibase 10.1103/PhysRevA.45.8185} {\bibfield  {journal}
  {\bibinfo  {journal} {Phys. Rev. A}\ }\textbf {\bibinfo {volume} {45}},\
  \bibinfo {pages} {8185--8189} (\bibinfo {year} {1992})}\BibitemShut {NoStop}%
\bibitem [{\citenamefont {Bliokh}\ \emph {et~al.}(2007)\citenamefont {Bliokh},
  \citenamefont {Bliokh}, \citenamefont {Savel’Ev},\ and\ \citenamefont
  {Nori}}]{bliokh2007semiclassical}%
  \BibitemOpen
  \bibfield  {author} {\bibinfo {author} {\bibfnamefont {Konstantin~Yu}\
  \bibnamefont {Bliokh}}, \bibinfo {author} {\bibfnamefont {Yury~P}\
  \bibnamefont {Bliokh}}, \bibinfo {author} {\bibfnamefont {Sergey}\
  \bibnamefont {Savel’Ev}}, \ and\ \bibinfo {author} {\bibfnamefont {Franco}\
  \bibnamefont {Nori}},\ }\bibfield  {title} {\enquote {\bibinfo {title}
  {Semiclassical dynamics of electron wave packet states with phase
  vortices},}\ }\href@noop {} {\bibfield  {journal} {\bibinfo  {journal}
  {Physical Review Letters}\ }\textbf {\bibinfo {volume} {99}},\ \bibinfo
  {pages} {190404} (\bibinfo {year} {2007})}\BibitemShut {NoStop}%
\bibitem [{\citenamefont {Uchida}\ and\ \citenamefont
  {Tonomura}(2010)}]{uchida2010generation}%
  \BibitemOpen
  \bibfield  {author} {\bibinfo {author} {\bibfnamefont {Masaya}\ \bibnamefont
  {Uchida}}\ and\ \bibinfo {author} {\bibfnamefont {Akira}\ \bibnamefont
  {Tonomura}},\ }\bibfield  {title} {\enquote {\bibinfo {title} {Generation of
  electron beams carrying orbital angular momentum},}\ }\href@noop {}
  {\bibfield  {journal} {\bibinfo  {journal} {nature}\ }\textbf {\bibinfo
  {volume} {464}},\ \bibinfo {pages} {737} (\bibinfo {year}
  {2010})}\BibitemShut {NoStop}%
\bibitem [{\citenamefont {McMorran}\ \emph {et~al.}(2011)\citenamefont
  {McMorran}, \citenamefont {Agrawal}, \citenamefont {Anderson}, \citenamefont
  {Herzing}, \citenamefont {Lezec}, \citenamefont {McClelland},\ and\
  \citenamefont {Unguris}}]{mcmorran2011electron}%
  \BibitemOpen
  \bibfield  {author} {\bibinfo {author} {\bibfnamefont {Benjamin~J}\
  \bibnamefont {McMorran}}, \bibinfo {author} {\bibfnamefont {Amit}\
  \bibnamefont {Agrawal}}, \bibinfo {author} {\bibfnamefont {Ian~M}\
  \bibnamefont {Anderson}}, \bibinfo {author} {\bibfnamefont {Andrew~A}\
  \bibnamefont {Herzing}}, \bibinfo {author} {\bibfnamefont {Henri~J}\
  \bibnamefont {Lezec}}, \bibinfo {author} {\bibfnamefont {Jabez~J}\
  \bibnamefont {McClelland}}, \ and\ \bibinfo {author} {\bibfnamefont {John}\
  \bibnamefont {Unguris}},\ }\bibfield  {title} {\enquote {\bibinfo {title}
  {Electron vortex beams with high quanta of orbital angular momentum},}\
  }\href@noop {} {\bibfield  {journal} {\bibinfo  {journal} {science}\ }\textbf
  {\bibinfo {volume} {331}},\ \bibinfo {pages} {192--195} (\bibinfo {year}
  {2011})}\BibitemShut {NoStop}%
\bibitem [{\citenamefont {Clark}\ \emph {et~al.}(2015)\citenamefont {Clark},
  \citenamefont {Barankov}, \citenamefont {Huber}, \citenamefont {Cory},\ and\
  \citenamefont {Pushin}}]{Dima2015}%
  \BibitemOpen
  \bibfield  {author} {\bibinfo {author} {\bibfnamefont {C.~W.}\ \bibnamefont
  {Clark}}, \bibinfo {author} {\bibfnamefont {R.}~\bibnamefont {Barankov}},
  \bibinfo {author} {\bibfnamefont {M.~G.}\ \bibnamefont {Huber}}, \bibinfo
  {author} {\bibfnamefont {D.~G.}\ \bibnamefont {Cory}}, \ and\ \bibinfo
  {author} {\bibfnamefont {D.~A.}\ \bibnamefont {Pushin}},\ }\bibfield  {title}
  {\enquote {\bibinfo {title} {Controlling neutron orbital angular momentum},}\
  }\href {\doibase http://dx.doi.org/10.1038/525462a} {\bibfield  {journal}
  {\bibinfo  {journal} {Nature}\ }\textbf {\bibinfo {volume} {525}},\ \bibinfo
  {pages} {504--506} (\bibinfo {year} {2015})}\BibitemShut {NoStop}%
\bibitem [{\citenamefont {Mair}\ \emph {et~al.}(2001)\citenamefont {Mair},
  \citenamefont {Vaziri}, \citenamefont {Weihs},\ and\ \citenamefont
  {Zeilinger}}]{mair2001entanglement}%
  \BibitemOpen
  \bibfield  {author} {\bibinfo {author} {\bibfnamefont {Alois}\ \bibnamefont
  {Mair}}, \bibinfo {author} {\bibfnamefont {Alipasha}\ \bibnamefont {Vaziri}},
  \bibinfo {author} {\bibfnamefont {Gregor}\ \bibnamefont {Weihs}}, \ and\
  \bibinfo {author} {\bibfnamefont {Anton}\ \bibnamefont {Zeilinger}},\
  }\bibfield  {title} {\enquote {\bibinfo {title} {Entanglement of the orbital
  angular momentum states of photons},}\ }\href@noop {} {\bibfield  {journal}
  {\bibinfo  {journal} {Nature}\ }\textbf {\bibinfo {volume} {412}},\ \bibinfo
  {pages} {313--316} (\bibinfo {year} {2001})}\BibitemShut {NoStop}%
\bibitem [{\citenamefont {Wang}\ \emph {et~al.}(2012)\citenamefont {Wang},
  \citenamefont {Yang}, \citenamefont {Fazal}, \citenamefont {Ahmed},
  \citenamefont {Yan}, \citenamefont {Huang}, \citenamefont {Ren},
  \citenamefont {Yue}, \citenamefont {Dolinar}, \citenamefont {Tur} \emph
  {et~al.}}]{wang2012terabit}%
  \BibitemOpen
  \bibfield  {author} {\bibinfo {author} {\bibfnamefont {Jian}\ \bibnamefont
  {Wang}}, \bibinfo {author} {\bibfnamefont {Jeng-Yuan}\ \bibnamefont {Yang}},
  \bibinfo {author} {\bibfnamefont {Irfan~M}\ \bibnamefont {Fazal}}, \bibinfo
  {author} {\bibfnamefont {Nisar}\ \bibnamefont {Ahmed}}, \bibinfo {author}
  {\bibfnamefont {Yan}\ \bibnamefont {Yan}}, \bibinfo {author} {\bibfnamefont
  {Hao}\ \bibnamefont {Huang}}, \bibinfo {author} {\bibfnamefont {Yongxiong}\
  \bibnamefont {Ren}}, \bibinfo {author} {\bibfnamefont {Yang}\ \bibnamefont
  {Yue}}, \bibinfo {author} {\bibfnamefont {Samuel}\ \bibnamefont {Dolinar}},
  \bibinfo {author} {\bibfnamefont {Moshe}\ \bibnamefont {Tur}},  \emph
  {et~al.},\ }\bibfield  {title} {\enquote {\bibinfo {title} {Terabit
  free-space data transmission employing orbital angular momentum
  multiplexing},}\ }\href@noop {} {\bibfield  {journal} {\bibinfo  {journal}
  {Nature Photonics}\ }\textbf {\bibinfo {volume} {6}},\ \bibinfo {pages}
  {488--496} (\bibinfo {year} {2012})}\BibitemShut {NoStop}%
\bibitem [{\citenamefont {{Andersen}}\ \emph {et~al.}(2006)\citenamefont
  {{Andersen}}, \citenamefont {{Ryu}}, \citenamefont {{Clad{\'e}}},
  \citenamefont {{Natarajan}}, \citenamefont {{Vaziri}}, \citenamefont
  {{Helmerson}},\ and\ \citenamefont {{Phillips}}}]{Andersen2006}%
  \BibitemOpen
  \bibfield  {author} {\bibinfo {author} {\bibfnamefont {M.~F.}\ \bibnamefont
  {{Andersen}}}, \bibinfo {author} {\bibfnamefont {C.}~\bibnamefont {{Ryu}}},
  \bibinfo {author} {\bibfnamefont {P.}~\bibnamefont {{Clad{\'e}}}}, \bibinfo
  {author} {\bibfnamefont {V.}~\bibnamefont {{Natarajan}}}, \bibinfo {author}
  {\bibfnamefont {A.}~\bibnamefont {{Vaziri}}}, \bibinfo {author}
  {\bibfnamefont {K.}~\bibnamefont {{Helmerson}}}, \ and\ \bibinfo {author}
  {\bibfnamefont {W.~D.}\ \bibnamefont {{Phillips}}},\ }\bibfield  {title}
  {\enquote {\bibinfo {title} {{Quantized Rotation of Atoms from Photons with
  Orbital Angular Momentum}},}\ }\href {\doibase 10.1103/PhysRevLett.97.170406}
  {\bibfield  {journal} {\bibinfo  {journal} {Physical Review Letters}\
  }\textbf {\bibinfo {volume} {97}},\ \bibinfo {eid} {170406} (\bibinfo {year}
  {2006})}\BibitemShut {NoStop}%
\bibitem [{\citenamefont {He}\ \emph {et~al.}(1995)\citenamefont {He},
  \citenamefont {Friese}, \citenamefont {Heckenberg},\ and\ \citenamefont
  {Rubinsztein-Dunlop}}]{he1995direct}%
  \BibitemOpen
  \bibfield  {author} {\bibinfo {author} {\bibfnamefont {H}~\bibnamefont {He}},
  \bibinfo {author} {\bibfnamefont {MEJ}\ \bibnamefont {Friese}}, \bibinfo
  {author} {\bibfnamefont {NR}~\bibnamefont {Heckenberg}}, \ and\ \bibinfo
  {author} {\bibfnamefont {H}~\bibnamefont {Rubinsztein-Dunlop}},\ }\bibfield
  {title} {\enquote {\bibinfo {title} {Direct observation of transfer of
  angular momentum to absorptive particles from a laser beam with a phase
  singularity},}\ }\href@noop {} {\bibfield  {journal} {\bibinfo  {journal}
  {Physical Review Letters}\ }\textbf {\bibinfo {volume} {75}},\ \bibinfo
  {pages} {826} (\bibinfo {year} {1995})}\BibitemShut {NoStop}%
\bibitem [{\citenamefont {Friese}\ \emph {et~al.}(1996)\citenamefont {Friese},
  \citenamefont {Enger}, \citenamefont {Rubinsztein-Dunlop},\ and\
  \citenamefont {Heckenberg}}]{friese1996optical}%
  \BibitemOpen
  \bibfield  {author} {\bibinfo {author} {\bibfnamefont {MEJ}\ \bibnamefont
  {Friese}}, \bibinfo {author} {\bibfnamefont {J}~\bibnamefont {Enger}},
  \bibinfo {author} {\bibfnamefont {H}~\bibnamefont {Rubinsztein-Dunlop}}, \
  and\ \bibinfo {author} {\bibfnamefont {Norman~R}\ \bibnamefont
  {Heckenberg}},\ }\bibfield  {title} {\enquote {\bibinfo {title} {Optical
  angular-momentum transfer to trapped absorbing particles},}\ }\href@noop {}
  {\bibfield  {journal} {\bibinfo  {journal} {Physical Review A}\ }\textbf
  {\bibinfo {volume} {54}},\ \bibinfo {pages} {1593} (\bibinfo {year}
  {1996})}\BibitemShut {NoStop}%
\bibitem [{\citenamefont {Brullot}\ \emph {et~al.}(2016)\citenamefont
  {Brullot}, \citenamefont {Vanbel}, \citenamefont {Swusten},\ and\
  \citenamefont {Verbiest}}]{brullot2016resolving}%
  \BibitemOpen
  \bibfield  {author} {\bibinfo {author} {\bibfnamefont {Ward}\ \bibnamefont
  {Brullot}}, \bibinfo {author} {\bibfnamefont {Maarten~K}\ \bibnamefont
  {Vanbel}}, \bibinfo {author} {\bibfnamefont {Tom}\ \bibnamefont {Swusten}}, \
  and\ \bibinfo {author} {\bibfnamefont {Thierry}\ \bibnamefont {Verbiest}},\
  }\bibfield  {title} {\enquote {\bibinfo {title} {Resolving enantiomers using
  the optical angular momentum of twisted light},}\ }\href@noop {} {\bibfield
  {journal} {\bibinfo  {journal} {Science advances}\ }\textbf {\bibinfo
  {volume} {2}},\ \bibinfo {pages} {e1501349} (\bibinfo {year}
  {2016})}\BibitemShut {NoStop}%
\bibitem [{\citenamefont {Simpson}\ \emph {et~al.}(1997)\citenamefont
  {Simpson}, \citenamefont {Dholakia}, \citenamefont {Allen},\ and\
  \citenamefont {Padgett}}]{Simpson:97}%
  \BibitemOpen
  \bibfield  {author} {\bibinfo {author} {\bibfnamefont {N.~B.}\ \bibnamefont
  {Simpson}}, \bibinfo {author} {\bibfnamefont {K.}~\bibnamefont {Dholakia}},
  \bibinfo {author} {\bibfnamefont {L.}~\bibnamefont {Allen}}, \ and\ \bibinfo
  {author} {\bibfnamefont {M.~J.}\ \bibnamefont {Padgett}},\ }\bibfield
  {title} {\enquote {\bibinfo {title} {Mechanical equivalence of spin and
  orbital angular momentum of light: an optical spanner},}\ }\href {\doibase
  10.1364/OL.22.000052} {\bibfield  {journal} {\bibinfo  {journal} {Opt.
  Lett.}\ }\textbf {\bibinfo {volume} {22}},\ \bibinfo {pages} {52--54}
  (\bibinfo {year} {1997})}\BibitemShut {NoStop}%
\bibitem [{\citenamefont {Zhang}\ \emph {et~al.}(2019)\citenamefont {Zhang},
  \citenamefont {Li}, \citenamefont {Zhao}, \citenamefont {Xie}, \citenamefont
  {Milione}, \citenamefont {Song}, \citenamefont {Liao}, \citenamefont {Liu},
  \citenamefont {Song}, \citenamefont {Pang}, \citenamefont {Willner},
  \citenamefont {Lynn}, \citenamefont {Bock}, \citenamefont {Tur},\ and\
  \citenamefont {Willner}}]{Zhang:19}%
  \BibitemOpen
  \bibfield  {author} {\bibinfo {author} {\bibfnamefont {Runzhou}\ \bibnamefont
  {Zhang}}, \bibinfo {author} {\bibfnamefont {Long}\ \bibnamefont {Li}},
  \bibinfo {author} {\bibfnamefont {Zhe}\ \bibnamefont {Zhao}}, \bibinfo
  {author} {\bibfnamefont {Guodong}\ \bibnamefont {Xie}}, \bibinfo {author}
  {\bibfnamefont {Giovanni}\ \bibnamefont {Milione}}, \bibinfo {author}
  {\bibfnamefont {Hao}\ \bibnamefont {Song}}, \bibinfo {author} {\bibfnamefont
  {Peicheng}\ \bibnamefont {Liao}}, \bibinfo {author} {\bibfnamefont {Cong}\
  \bibnamefont {Liu}}, \bibinfo {author} {\bibfnamefont {Haoqian}\ \bibnamefont
  {Song}}, \bibinfo {author} {\bibfnamefont {Kai}\ \bibnamefont {Pang}},
  \bibinfo {author} {\bibfnamefont {Ari~N.}\ \bibnamefont {Willner}}, \bibinfo
  {author} {\bibfnamefont {Brittany}\ \bibnamefont {Lynn}}, \bibinfo {author}
  {\bibfnamefont {Robert}\ \bibnamefont {Bock}}, \bibinfo {author}
  {\bibfnamefont {Moshe}\ \bibnamefont {Tur}}, \ and\ \bibinfo {author}
  {\bibfnamefont {Alan~E.}\ \bibnamefont {Willner}},\ }\bibfield  {title}
  {\enquote {\bibinfo {title} {Coherent optical wireless communication link
  employing orbital angular momentum multiplexing in a ballistic and diffusive
  scattering medium},}\ }\href {\doibase 10.1364/OL.44.000691} {\bibfield
  {journal} {\bibinfo  {journal} {Opt. Lett.}\ }\textbf {\bibinfo {volume}
  {44}},\ \bibinfo {pages} {691--694} (\bibinfo {year} {2019})}\BibitemShut
  {NoStop}%
\bibitem [{\citenamefont {Grillo}\ \emph {et~al.}(2017)\citenamefont {Grillo},
  \citenamefont {Harvey}, \citenamefont {Venturi}, \citenamefont {Pierce},
  \citenamefont {Balboni}, \citenamefont {Bouchard}, \citenamefont {Gazzadi},
  \citenamefont {Frabboni}, \citenamefont {Tavabi}, \citenamefont {Li} \emph
  {et~al.}}]{grillo2017observation}%
  \BibitemOpen
  \bibfield  {author} {\bibinfo {author} {\bibfnamefont {Vincenzo}\
  \bibnamefont {Grillo}}, \bibinfo {author} {\bibfnamefont {Tyler~R}\
  \bibnamefont {Harvey}}, \bibinfo {author} {\bibfnamefont {Federico}\
  \bibnamefont {Venturi}}, \bibinfo {author} {\bibfnamefont {Jordan~S}\
  \bibnamefont {Pierce}}, \bibinfo {author} {\bibfnamefont {Roberto}\
  \bibnamefont {Balboni}}, \bibinfo {author} {\bibfnamefont {Fr{\'e}d{\'e}ric}\
  \bibnamefont {Bouchard}}, \bibinfo {author} {\bibfnamefont {Gian~Carlo}\
  \bibnamefont {Gazzadi}}, \bibinfo {author} {\bibfnamefont {Stefano}\
  \bibnamefont {Frabboni}}, \bibinfo {author} {\bibfnamefont {Amir~H}\
  \bibnamefont {Tavabi}}, \bibinfo {author} {\bibfnamefont {Zi-An}\
  \bibnamefont {Li}},  \emph {et~al.},\ }\bibfield  {title} {\enquote {\bibinfo
  {title} {Observation of nanoscale magnetic fields using twisted electron
  beams},}\ }\href@noop {} {\bibfield  {journal} {\bibinfo  {journal} {Nature
  communications}\ }\textbf {\bibinfo {volume} {8}},\ \bibinfo {pages} {689}
  (\bibinfo {year} {2017})}\BibitemShut {NoStop}%
\bibitem [{\citenamefont {B{\'e}ch{\'e}}\ \emph {et~al.}(2014)\citenamefont
  {B{\'e}ch{\'e}}, \citenamefont {Van~Boxem}, \citenamefont {Van~Tendeloo},\
  and\ \citenamefont {Verbeeck}}]{beche2014magnetic}%
  \BibitemOpen
  \bibfield  {author} {\bibinfo {author} {\bibfnamefont {Armand}\ \bibnamefont
  {B{\'e}ch{\'e}}}, \bibinfo {author} {\bibfnamefont {Ruben}\ \bibnamefont
  {Van~Boxem}}, \bibinfo {author} {\bibfnamefont {Gustaaf}\ \bibnamefont
  {Van~Tendeloo}}, \ and\ \bibinfo {author} {\bibfnamefont {Jo}~\bibnamefont
  {Verbeeck}},\ }\bibfield  {title} {\enquote {\bibinfo {title} {Magnetic
  monopole field exposed by electrons},}\ }\href@noop {} {\bibfield  {journal}
  {\bibinfo  {journal} {Nature Physics}\ }\textbf {\bibinfo {volume} {10}},\
  \bibinfo {pages} {26} (\bibinfo {year} {2014})}\BibitemShut {NoStop}%
\bibitem [{\citenamefont {Sarenac}\ \emph {et~al.}(2016)\citenamefont
  {Sarenac}, \citenamefont {Huber}, \citenamefont {Heacock}, \citenamefont
  {Arif}, \citenamefont {Clark}, \citenamefont {Cory}, \citenamefont {Shahi},\
  and\ \citenamefont {Pushin}}]{holography}%
  \BibitemOpen
  \bibfield  {author} {\bibinfo {author} {\bibfnamefont {D.}~\bibnamefont
  {Sarenac}}, \bibinfo {author} {\bibfnamefont {M.~G.}\ \bibnamefont {Huber}},
  \bibinfo {author} {\bibfnamefont {B.}~\bibnamefont {Heacock}}, \bibinfo
  {author} {\bibfnamefont {M.}~\bibnamefont {Arif}}, \bibinfo {author}
  {\bibfnamefont {C.~W.}\ \bibnamefont {Clark}}, \bibinfo {author}
  {\bibfnamefont {D.~G.}\ \bibnamefont {Cory}}, \bibinfo {author}
  {\bibfnamefont {C.~B.}\ \bibnamefont {Shahi}}, \ and\ \bibinfo {author}
  {\bibfnamefont {D.~A.}\ \bibnamefont {Pushin}},\ }\bibfield  {title}
  {\enquote {\bibinfo {title} {Holography with a neutron interferometer},}\
  }\href@noop {} {\bibfield  {journal} {\bibinfo  {journal} {Optics Express}\
  }\textbf {\bibinfo {volume} {24}},\ \bibinfo {pages} {22528} (\bibinfo {year}
  {2016})}\BibitemShut {NoStop}%
\bibitem [{\citenamefont {Afanasev}\ \emph {et~al.}(2019)\citenamefont
  {Afanasev}, \citenamefont {Karlovets},\ and\ \citenamefont
  {Serbo}}]{SchwingerOam}%
  \BibitemOpen
  \bibfield  {author} {\bibinfo {author} {\bibfnamefont {V.~Andrei}\
  \bibnamefont {Afanasev}}, \bibinfo {author} {\bibfnamefont {D.V.}\
  \bibnamefont {Karlovets}}, \ and\ \bibinfo {author} {\bibfnamefont {V.G.}\
  \bibnamefont {Serbo}},\ }\bibfield  {title} {\enquote {\bibinfo {title} {The
  schwinger scattering of twisted neutrons by nuclei},}\ }\href@noop {}
  {\bibfield  {journal} {\bibinfo  {journal} {arXiv preprint arXiv:1903.12245}\
  } (\bibinfo {year} {2019})}\BibitemShut {NoStop}%
\bibitem [{\citenamefont {Larocque}\ \emph {et~al.}(2018)\citenamefont
  {Larocque}, \citenamefont {Kaminer}, \citenamefont {Grillo}, \citenamefont
  {Boyd},\ and\ \citenamefont {Karimi}}]{larocque2018twisting}%
  \BibitemOpen
  \bibfield  {author} {\bibinfo {author} {\bibfnamefont {Hugo}\ \bibnamefont
  {Larocque}}, \bibinfo {author} {\bibfnamefont {Ido}\ \bibnamefont {Kaminer}},
  \bibinfo {author} {\bibfnamefont {Vincenzo}\ \bibnamefont {Grillo}}, \bibinfo
  {author} {\bibfnamefont {Robert~W}\ \bibnamefont {Boyd}}, \ and\ \bibinfo
  {author} {\bibfnamefont {Ebrahim}\ \bibnamefont {Karimi}},\ }\bibfield
  {title} {\enquote {\bibinfo {title} {Twisting neutrons may reveal their
  internal structure},}\ }\href@noop {} {\bibfield  {journal} {\bibinfo
  {journal} {Nature Physics}\ }\textbf {\bibinfo {volume} {14}},\ \bibinfo
  {pages} {1} (\bibinfo {year} {2018})}\BibitemShut {NoStop}%
\bibitem [{\citenamefont {Maurer}\ \emph {et~al.}(2007)\citenamefont {Maurer},
  \citenamefont {Jesacher}, \citenamefont {F{\"u}rhapter}, \citenamefont
  {Bernet},\ and\ \citenamefont {Ritsch-Marte}}]{maurer2007tailoring}%
  \BibitemOpen
  \bibfield  {author} {\bibinfo {author} {\bibfnamefont {Christian}\
  \bibnamefont {Maurer}}, \bibinfo {author} {\bibfnamefont {Alexander}\
  \bibnamefont {Jesacher}}, \bibinfo {author} {\bibfnamefont {Severin}\
  \bibnamefont {F{\"u}rhapter}}, \bibinfo {author} {\bibfnamefont {Stefan}\
  \bibnamefont {Bernet}}, \ and\ \bibinfo {author} {\bibfnamefont {Monika}\
  \bibnamefont {Ritsch-Marte}},\ }\bibfield  {title} {\enquote {\bibinfo
  {title} {Tailoring of arbitrary optical vector beams},}\ }\href@noop {}
  {\bibfield  {journal} {\bibinfo  {journal} {New Journal of Physics}\ }\textbf
  {\bibinfo {volume} {9}},\ \bibinfo {pages} {78} (\bibinfo {year}
  {2007})}\BibitemShut {NoStop}%
\bibitem [{\citenamefont {Marrucci}\ \emph {et~al.}(2006)\citenamefont
  {Marrucci}, \citenamefont {Manzo},\ and\ \citenamefont {Paparo}}]{qplate}%
  \BibitemOpen
  \bibfield  {author} {\bibinfo {author} {\bibfnamefont {L.}~\bibnamefont
  {Marrucci}}, \bibinfo {author} {\bibfnamefont {C.}~\bibnamefont {Manzo}}, \
  and\ \bibinfo {author} {\bibfnamefont {D.}~\bibnamefont {Paparo}},\
  }\bibfield  {title} {\enquote {\bibinfo {title} {Optical spin-to-orbital
  angular momentum conversion in inhomogeneous anisotropic media},}\ }\href
  {\doibase 10.1103/PhysRevLett.96.163905} {\bibfield  {journal} {\bibinfo
  {journal} {Phys. Rev. Lett.}\ }\textbf {\bibinfo {volume} {96}},\ \bibinfo
  {pages} {163905} (\bibinfo {year} {2006})}\BibitemShut {NoStop}%
\bibitem [{\citenamefont {Karimi}\ \emph {et~al.}(2012)\citenamefont {Karimi},
  \citenamefont {Marrucci}, \citenamefont {Grillo},\ and\ \citenamefont
  {Santamato}}]{karimi2012spin}%
  \BibitemOpen
  \bibfield  {author} {\bibinfo {author} {\bibfnamefont {Ebrahim}\ \bibnamefont
  {Karimi}}, \bibinfo {author} {\bibfnamefont {Lorenzo}\ \bibnamefont
  {Marrucci}}, \bibinfo {author} {\bibfnamefont {Vincenzo}\ \bibnamefont
  {Grillo}}, \ and\ \bibinfo {author} {\bibfnamefont {Enrico}\ \bibnamefont
  {Santamato}},\ }\bibfield  {title} {\enquote {\bibinfo {title}
  {Spin-to-orbital angular momentum conversion and spin-polarization filtering
  in electron beams},}\ }\href@noop {} {\bibfield  {journal} {\bibinfo
  {journal} {Physical review letters}\ }\textbf {\bibinfo {volume} {108}},\
  \bibinfo {pages} {044801} (\bibinfo {year} {2012})}\BibitemShut {NoStop}%
\bibitem [{\citenamefont {Nsofini}\ \emph {et~al.}(2016)\citenamefont
  {Nsofini}, \citenamefont {Sarenac}, \citenamefont {Wood}, \citenamefont
  {Cory}, \citenamefont {Arif}, \citenamefont {Clark}, \citenamefont {Huber},\
  and\ \citenamefont {Pushin}}]{spinorbit}%
  \BibitemOpen
  \bibfield  {author} {\bibinfo {author} {\bibfnamefont {Joachim}\ \bibnamefont
  {Nsofini}}, \bibinfo {author} {\bibfnamefont {Dusan}\ \bibnamefont
  {Sarenac}}, \bibinfo {author} {\bibfnamefont {Christopher~J.}\ \bibnamefont
  {Wood}}, \bibinfo {author} {\bibfnamefont {David~G.}\ \bibnamefont {Cory}},
  \bibinfo {author} {\bibfnamefont {Muhammad}\ \bibnamefont {Arif}}, \bibinfo
  {author} {\bibfnamefont {Charles~W.}\ \bibnamefont {Clark}}, \bibinfo
  {author} {\bibfnamefont {Michael~G.}\ \bibnamefont {Huber}}, \ and\ \bibinfo
  {author} {\bibfnamefont {Dmitry~A.}\ \bibnamefont {Pushin}},\ }\bibfield
  {title} {\enquote {\bibinfo {title} {Spin-orbit states of neutron wave
  packets},}\ }\href {\doibase 10.1103/PhysRevA.94.013605} {\bibfield
  {journal} {\bibinfo  {journal} {Phys. Rev. A}\ }\textbf {\bibinfo {volume}
  {94}},\ \bibinfo {pages} {013605} (\bibinfo {year} {2016})}\BibitemShut
  {NoStop}%
\bibitem [{\citenamefont {Marrucci}\ \emph {et~al.}(2011)\citenamefont
  {Marrucci}, \citenamefont {Karimi}, \citenamefont {Slussarenko},
  \citenamefont {Piccirillo}, \citenamefont {Santamato}, \citenamefont
  {Nagali},\ and\ \citenamefont {Sciarrino}}]{marrucci2011spin}%
  \BibitemOpen
  \bibfield  {author} {\bibinfo {author} {\bibfnamefont {Lorenzo}\ \bibnamefont
  {Marrucci}}, \bibinfo {author} {\bibfnamefont {Ebrahim}\ \bibnamefont
  {Karimi}}, \bibinfo {author} {\bibfnamefont {Sergei}\ \bibnamefont
  {Slussarenko}}, \bibinfo {author} {\bibfnamefont {Bruno}\ \bibnamefont
  {Piccirillo}}, \bibinfo {author} {\bibfnamefont {Enrico}\ \bibnamefont
  {Santamato}}, \bibinfo {author} {\bibfnamefont {Eleonora}\ \bibnamefont
  {Nagali}}, \ and\ \bibinfo {author} {\bibfnamefont {Fabio}\ \bibnamefont
  {Sciarrino}},\ }\bibfield  {title} {\enquote {\bibinfo {title}
  {Spin-to-orbital conversion of the angular momentum of light and its
  classical and quantum applications},}\ }\href@noop {} {\bibfield  {journal}
  {\bibinfo  {journal} {Journal of Optics}\ }\textbf {\bibinfo {volume} {13}},\
  \bibinfo {pages} {064001} (\bibinfo {year} {2011})}\BibitemShut {NoStop}%
\bibitem [{\citenamefont {Milione}\ \emph {et~al.}(2015)\citenamefont
  {Milione}, \citenamefont {Lavery}, \citenamefont {Huang}, \citenamefont
  {Ren}, \citenamefont {Xie}, \citenamefont {Nguyen}, \citenamefont {Karimi},
  \citenamefont {Marrucci}, \citenamefont {Nolan}, \citenamefont {Alfano} \emph
  {et~al.}}]{milione20154}%
  \BibitemOpen
  \bibfield  {author} {\bibinfo {author} {\bibfnamefont {Giovanni}\
  \bibnamefont {Milione}}, \bibinfo {author} {\bibfnamefont {Martin~PJ}\
  \bibnamefont {Lavery}}, \bibinfo {author} {\bibfnamefont {Hao}\ \bibnamefont
  {Huang}}, \bibinfo {author} {\bibfnamefont {Yongxiong}\ \bibnamefont {Ren}},
  \bibinfo {author} {\bibfnamefont {Guodong}\ \bibnamefont {Xie}}, \bibinfo
  {author} {\bibfnamefont {Thien~An}\ \bibnamefont {Nguyen}}, \bibinfo {author}
  {\bibfnamefont {Ebrahim}\ \bibnamefont {Karimi}}, \bibinfo {author}
  {\bibfnamefont {Lorenzo}\ \bibnamefont {Marrucci}}, \bibinfo {author}
  {\bibfnamefont {Daniel~A}\ \bibnamefont {Nolan}}, \bibinfo {author}
  {\bibfnamefont {Robert~R}\ \bibnamefont {Alfano}},  \emph {et~al.},\
  }\bibfield  {title} {\enquote {\bibinfo {title} {4$\times$ 20 gbit/s mode
  division multiplexing over free space using vector modes and a q-plate mode
  (de) multiplexer},}\ }\href@noop {} {\bibfield  {journal} {\bibinfo
  {journal} {Optics letters}\ }\textbf {\bibinfo {volume} {40}},\ \bibinfo
  {pages} {1980--1983} (\bibinfo {year} {2015})}\BibitemShut {NoStop}%
\bibitem [{\citenamefont {Schmiegelow}\ \emph {et~al.}(2016)\citenamefont
  {Schmiegelow}, \citenamefont {Schulz}, \citenamefont {Kaufmann},
  \citenamefont {Ruster}, \citenamefont {Poschinger},\ and\ \citenamefont
  {Schmidt-Kaler}}]{schmiegelow2016transfer}%
  \BibitemOpen
  \bibfield  {author} {\bibinfo {author} {\bibfnamefont {Christian~T}\
  \bibnamefont {Schmiegelow}}, \bibinfo {author} {\bibfnamefont {Jonas}\
  \bibnamefont {Schulz}}, \bibinfo {author} {\bibfnamefont {Henning}\
  \bibnamefont {Kaufmann}}, \bibinfo {author} {\bibfnamefont {Thomas}\
  \bibnamefont {Ruster}}, \bibinfo {author} {\bibfnamefont {Ulrich~G}\
  \bibnamefont {Poschinger}}, \ and\ \bibinfo {author} {\bibfnamefont
  {Ferdinand}\ \bibnamefont {Schmidt-Kaler}},\ }\bibfield  {title} {\enquote
  {\bibinfo {title} {Transfer of optical orbital angular momentum to a bound
  electron},}\ }\href@noop {} {\bibfield  {journal} {\bibinfo  {journal}
  {Nature communications}\ }\textbf {\bibinfo {volume} {7}} (\bibinfo {year}
  {2016})}\BibitemShut {NoStop}%
\bibitem [{\citenamefont {Vallone}\ \emph {et~al.}(2014)\citenamefont
  {Vallone}, \citenamefont {D'Ambrosio}, \citenamefont {Sponselli},
  \citenamefont {Slussarenko}, \citenamefont {Marrucci}, \citenamefont
  {Sciarrino},\ and\ \citenamefont {Villoresi}}]{vallone2014freespace}%
  \BibitemOpen
  \bibfield  {author} {\bibinfo {author} {\bibfnamefont {Giuseppe}\
  \bibnamefont {Vallone}}, \bibinfo {author} {\bibfnamefont {Vincenzo}\
  \bibnamefont {D'Ambrosio}}, \bibinfo {author} {\bibfnamefont {Anna}\
  \bibnamefont {Sponselli}}, \bibinfo {author} {\bibfnamefont {Sergei}\
  \bibnamefont {Slussarenko}}, \bibinfo {author} {\bibfnamefont {Lorenzo}\
  \bibnamefont {Marrucci}}, \bibinfo {author} {\bibfnamefont {Fabio}\
  \bibnamefont {Sciarrino}}, \ and\ \bibinfo {author} {\bibfnamefont {Paolo}\
  \bibnamefont {Villoresi}},\ }\bibfield  {title} {\enquote {\bibinfo {title}
  {Free-space quantum key distribution by rotation-invariant twisted
  photons},}\ }\href {\doibase 10.1103/PhysRevLett.113.060503} {\bibfield
  {journal} {\bibinfo  {journal} {Phys. Rev. Lett.}\ }\textbf {\bibinfo
  {volume} {113}},\ \bibinfo {pages} {060503} (\bibinfo {year}
  {2014})}\BibitemShut {NoStop}%
\bibitem [{\citenamefont {Karimi}\ \emph {et~al.}(2014)\citenamefont {Karimi},
  \citenamefont {Grillo}, \citenamefont {Boyd},\ and\ \citenamefont
  {Santamato}}]{karimi2014generation}%
  \BibitemOpen
  \bibfield  {author} {\bibinfo {author} {\bibfnamefont {Ebrahim}\ \bibnamefont
  {Karimi}}, \bibinfo {author} {\bibfnamefont {Vincenzo}\ \bibnamefont
  {Grillo}}, \bibinfo {author} {\bibfnamefont {Robert~W}\ \bibnamefont {Boyd}},
  \ and\ \bibinfo {author} {\bibfnamefont {Enrico}\ \bibnamefont {Santamato}},\
  }\bibfield  {title} {\enquote {\bibinfo {title} {Generation of a
  spin-polarized electron beam by multipole magnetic fields},}\ }\href@noop {}
  {\bibfield  {journal} {\bibinfo  {journal} {Ultramicroscopy}\ }\textbf
  {\bibinfo {volume} {138}},\ \bibinfo {pages} {22--27} (\bibinfo {year}
  {2014})}\BibitemShut {NoStop}%
\bibitem [{\citenamefont {Sarenac}\ \emph
  {et~al.}(2018{\natexlab{a}})\citenamefont {Sarenac}, \citenamefont {Cory},
  \citenamefont {Nsofini}, \citenamefont {Hincks}, \citenamefont {Miguel},
  \citenamefont {Arif}, \citenamefont {Clark}, \citenamefont {Huber},\ and\
  \citenamefont {Pushin}}]{sarenac2018generation}%
  \BibitemOpen
  \bibfield  {author} {\bibinfo {author} {\bibfnamefont {D}~\bibnamefont
  {Sarenac}}, \bibinfo {author} {\bibfnamefont {DG}~\bibnamefont {Cory}},
  \bibinfo {author} {\bibfnamefont {J}~\bibnamefont {Nsofini}}, \bibinfo
  {author} {\bibfnamefont {I}~\bibnamefont {Hincks}}, \bibinfo {author}
  {\bibfnamefont {P}~\bibnamefont {Miguel}}, \bibinfo {author} {\bibfnamefont
  {M}~\bibnamefont {Arif}}, \bibinfo {author} {\bibfnamefont {Charles~W}\
  \bibnamefont {Clark}}, \bibinfo {author} {\bibfnamefont {MG}~\bibnamefont
  {Huber}}, \ and\ \bibinfo {author} {\bibfnamefont {DA}~\bibnamefont
  {Pushin}},\ }\bibfield  {title} {\enquote {\bibinfo {title} {Generation of a
  lattice of spin-orbit beams via coherent averaging},}\ }\href@noop {}
  {\bibfield  {journal} {\bibinfo  {journal} {Physical review letters}\
  }\textbf {\bibinfo {volume} {121}},\ \bibinfo {pages} {183602} (\bibinfo
  {year} {2018}{\natexlab{a}})}\BibitemShut {NoStop}%
\bibitem [{\citenamefont {Sarenac}\ \emph
  {et~al.}(2018{\natexlab{b}})\citenamefont {Sarenac}, \citenamefont {Nsofini},
  \citenamefont {Hincks}, \citenamefont {Arif}, \citenamefont {Clark},
  \citenamefont {Cory}, \citenamefont {Huber},\ and\ \citenamefont
  {Pushin}}]{methods}%
  \BibitemOpen
  \bibfield  {author} {\bibinfo {author} {\bibfnamefont {D}~\bibnamefont
  {Sarenac}}, \bibinfo {author} {\bibfnamefont {J}~\bibnamefont {Nsofini}},
  \bibinfo {author} {\bibfnamefont {I}~\bibnamefont {Hincks}}, \bibinfo
  {author} {\bibfnamefont {M}~\bibnamefont {Arif}}, \bibinfo {author}
  {\bibfnamefont {Charles~W}\ \bibnamefont {Clark}}, \bibinfo {author}
  {\bibfnamefont {DG}~\bibnamefont {Cory}}, \bibinfo {author} {\bibfnamefont
  {MG}~\bibnamefont {Huber}}, \ and\ \bibinfo {author} {\bibfnamefont
  {DA}~\bibnamefont {Pushin}},\ }\bibfield  {title} {\enquote {\bibinfo {title}
  {Methods for preparation and detection of neutron spin-orbit states},}\
  }\href@noop {} {\bibfield  {journal} {\bibinfo  {journal} {New Journal of
  Physics}\ }\textbf {\bibinfo {volume} {20}},\ \bibinfo {pages} {103012}
  (\bibinfo {year} {2018}{\natexlab{b}})}\BibitemShut {NoStop}%
\bibitem [{\citenamefont {{Tsesses}}\ \emph {et~al.}(2018)\citenamefont
  {{Tsesses}}, \citenamefont {{Ostrovsky}}, \citenamefont {{Cohen}},
  \citenamefont {{Gjonaj}}, \citenamefont {{Lindner}},\ and\ \citenamefont
  {{Bartal}}}]{2018Sci...361..993T}%
  \BibitemOpen
  \bibfield  {author} {\bibinfo {author} {\bibfnamefont {S.}~\bibnamefont
  {{Tsesses}}}, \bibinfo {author} {\bibfnamefont {E.}~\bibnamefont
  {{Ostrovsky}}}, \bibinfo {author} {\bibfnamefont {K.}~\bibnamefont
  {{Cohen}}}, \bibinfo {author} {\bibfnamefont {B.}~\bibnamefont {{Gjonaj}}},
  \bibinfo {author} {\bibfnamefont {N.~H.}\ \bibnamefont {{Lindner}}}, \ and\
  \bibinfo {author} {\bibfnamefont {G.}~\bibnamefont {{Bartal}}},\ }\bibfield
  {title} {\enquote {\bibinfo {title} {{Optical skyrmion lattice in evanescent
  electromagnetic fields}},}\ }\href {\doibase 10.1126/science.aau0227}
  {\bibfield  {journal} {\bibinfo  {journal} {Science}\ }\textbf {\bibinfo
  {volume} {361}},\ \bibinfo {pages} {993--996} (\bibinfo {year} {2018})},\
  \Eprint {http://arxiv.org/abs/1805.11839} {arXiv:1805.11839 [physics.optics]}
  \BibitemShut {NoStop}%
\bibitem [{pha()}]{phades}%
  \BibitemOpen
  \href@noop {} {}\bibinfo {note}
  {{h}ttps://ncnr.nist.gov/instruments/instdev.html}\BibitemShut {NoStop}%
\bibitem [{\citenamefont {Dietze}\ \emph {et~al.}(1996)\citenamefont {Dietze},
  \citenamefont {Felber}, \citenamefont {Raum},\ and\ \citenamefont
  {Rausch}}]{dietze1996intensified}%
  \BibitemOpen
  \bibfield  {author} {\bibinfo {author} {\bibfnamefont {M}~\bibnamefont
  {Dietze}}, \bibinfo {author} {\bibfnamefont {J}~\bibnamefont {Felber}},
  \bibinfo {author} {\bibfnamefont {K}~\bibnamefont {Raum}}, \ and\ \bibinfo
  {author} {\bibfnamefont {C}~\bibnamefont {Rausch}},\ }\bibfield  {title}
  {\enquote {\bibinfo {title} {Intensified ccds as position sensitive neutron
  detectors},}\ }\href@noop {} {\bibfield  {journal} {\bibinfo  {journal}
  {Nuclear Instruments and Methods in Physics Research Section A: Accelerators,
  Spectrometers, Detectors and Associated Equipment}\ }\textbf {\bibinfo
  {volume} {377}},\ \bibinfo {pages} {320--324} (\bibinfo {year}
  {1996})}\BibitemShut {NoStop}%
\bibitem [{\citenamefont {Chen}\ \emph
  {et~al.}(2014{\natexlab{a}})\citenamefont {Chen}, \citenamefont {Gentile},
  \citenamefont {Erwin}, \citenamefont {Watson}, \citenamefont {Ye},
  \citenamefont {Krycka},\ and\ \citenamefont {Maranville}}]{chen20143he}%
  \BibitemOpen
  \bibfield  {author} {\bibinfo {author} {\bibfnamefont {WangChun}\
  \bibnamefont {Chen}}, \bibinfo {author} {\bibfnamefont {Thomas~R}\
  \bibnamefont {Gentile}}, \bibinfo {author} {\bibfnamefont {R}~\bibnamefont
  {Erwin}}, \bibinfo {author} {\bibfnamefont {Shannon}\ \bibnamefont {Watson}},
  \bibinfo {author} {\bibfnamefont {Q}~\bibnamefont {Ye}}, \bibinfo {author}
  {\bibfnamefont {Kathryn~L}\ \bibnamefont {Krycka}}, \ and\ \bibinfo {author}
  {\bibfnamefont {Brian~B}\ \bibnamefont {Maranville}},\ }\bibfield  {title}
  {\enquote {\bibinfo {title} {3he spin filter based polarized neutron
  capability at the nist center for neutron research},}\ }in\ \href@noop {}
  {\emph {\bibinfo {booktitle} {Journal of Physics: Conference Series}}},\
  Vol.\ \bibinfo {volume} {528}\ (\bibinfo {organization} {IOP Publishing},\
  \bibinfo {year} {2014})\ p.\ \bibinfo {pages} {012014}\BibitemShut {NoStop}%
\bibitem [{\citenamefont {Chen}\ \emph
  {et~al.}(2014{\natexlab{b}})\citenamefont {Chen}, \citenamefont {Gentile},
  \citenamefont {Ye}, \citenamefont {Walker},\ and\ \citenamefont
  {Babcock}}]{chen2014limits}%
  \BibitemOpen
  \bibfield  {author} {\bibinfo {author} {\bibfnamefont {WC}~\bibnamefont
  {Chen}}, \bibinfo {author} {\bibfnamefont {TR}~\bibnamefont {Gentile}},
  \bibinfo {author} {\bibfnamefont {Q}~\bibnamefont {Ye}}, \bibinfo {author}
  {\bibfnamefont {TG}~\bibnamefont {Walker}}, \ and\ \bibinfo {author}
  {\bibfnamefont {E}~\bibnamefont {Babcock}},\ }\bibfield  {title} {\enquote
  {\bibinfo {title} {On the limits of spin-exchange optical pumping of 3he},}\
  }\href@noop {} {\bibfield  {journal} {\bibinfo  {journal} {Journal of applied
  physics}\ }\textbf {\bibinfo {volume} {116}},\ \bibinfo {pages} {014903}
  (\bibinfo {year} {2014}{\natexlab{b}})}\BibitemShut {NoStop}%
\bibitem [{\citenamefont {Abragam}\ and\ \citenamefont
  {Abragam}(1961)}]{abragam1961principles}%
  \BibitemOpen
  \bibfield  {author} {\bibinfo {author} {\bibfnamefont {Anatole}\ \bibnamefont
  {Abragam}}\ and\ \bibinfo {author} {\bibfnamefont {Anatole}\ \bibnamefont
  {Abragam}},\ }\href@noop {} {\emph {\bibinfo {title} {The principles of
  nuclear magnetism}}},\ \bibinfo {number} {32}\ (\bibinfo  {publisher} {Oxford
  university press},\ \bibinfo {year} {1961})\BibitemShut {NoStop}%
\bibitem [{\citenamefont {Li}\ \emph {et~al.}(2017)\citenamefont {Li},
  \citenamefont {Feng}, \citenamefont {Thaler}, \citenamefont {Parnell},
  \citenamefont {Hamilton}, \citenamefont {Crow}, \citenamefont {Yang},
  \citenamefont {Jones}, \citenamefont {Bai}, \citenamefont {Matsuda} \emph
  {et~al.}}]{li2017high}%
  \BibitemOpen
  \bibfield  {author} {\bibinfo {author} {\bibfnamefont {Fankang}\ \bibnamefont
  {Li}}, \bibinfo {author} {\bibfnamefont {Hao}\ \bibnamefont {Feng}}, \bibinfo
  {author} {\bibfnamefont {Alexander~N}\ \bibnamefont {Thaler}}, \bibinfo
  {author} {\bibfnamefont {Steven~R}\ \bibnamefont {Parnell}}, \bibinfo
  {author} {\bibfnamefont {William~A}\ \bibnamefont {Hamilton}}, \bibinfo
  {author} {\bibfnamefont {Lowell}\ \bibnamefont {Crow}}, \bibinfo {author}
  {\bibfnamefont {Wencao}\ \bibnamefont {Yang}}, \bibinfo {author}
  {\bibfnamefont {Amy~B}\ \bibnamefont {Jones}}, \bibinfo {author}
  {\bibfnamefont {Hongyu}\ \bibnamefont {Bai}}, \bibinfo {author}
  {\bibfnamefont {Masaaki}\ \bibnamefont {Matsuda}},  \emph {et~al.},\
  }\bibfield  {title} {\enquote {\bibinfo {title} {High resolution neutron
  larmor diffraction using superconducting magnetic wollaston prisms},}\
  }\href@noop {} {\bibfield  {journal} {\bibinfo  {journal} {Scientific
  reports}\ }\textbf {\bibinfo {volume} {7}},\ \bibinfo {pages} {865} (\bibinfo
  {year} {2017})}\BibitemShut {NoStop}%
\bibitem [{\citenamefont {Bouwman}\ \emph {et~al.}(2008)\citenamefont
  {Bouwman}, \citenamefont {Plomp}, \citenamefont {De~Haan}, \citenamefont
  {Kraan}, \citenamefont {van Well}, \citenamefont {Habicht}, \citenamefont
  {Keller},\ and\ \citenamefont {Rekveldt}}]{bouwman2008real}%
  \BibitemOpen
  \bibfield  {author} {\bibinfo {author} {\bibfnamefont {Wim~G}\ \bibnamefont
  {Bouwman}}, \bibinfo {author} {\bibfnamefont {Jeroen}\ \bibnamefont {Plomp}},
  \bibinfo {author} {\bibfnamefont {Victor~O}\ \bibnamefont {De~Haan}},
  \bibinfo {author} {\bibfnamefont {Wicher~H}\ \bibnamefont {Kraan}}, \bibinfo
  {author} {\bibfnamefont {Ad~A}\ \bibnamefont {van Well}}, \bibinfo {author}
  {\bibfnamefont {Klaus}\ \bibnamefont {Habicht}}, \bibinfo {author}
  {\bibfnamefont {Thomas}\ \bibnamefont {Keller}}, \ and\ \bibinfo {author}
  {\bibfnamefont {M~Theo}\ \bibnamefont {Rekveldt}},\ }\bibfield  {title}
  {\enquote {\bibinfo {title} {Real-space neutron scattering methods},}\
  }\href@noop {} {\bibfield  {journal} {\bibinfo  {journal} {Nuclear
  Instruments and Methods in Physics Research Section A: Accelerators,
  Spectrometers, Detectors and Associated Equipment}\ }\textbf {\bibinfo
  {volume} {586}},\ \bibinfo {pages} {9--14} (\bibinfo {year}
  {2008})}\BibitemShut {NoStop}%
\bibitem [{\citenamefont {Mohr}\ \emph {et~al.}(2016)\citenamefont {Mohr},
  \citenamefont {Newell},\ and\ \citenamefont {Taylor}}]{codata}%
  \BibitemOpen
  \bibfield  {author} {\bibinfo {author} {\bibfnamefont {Peter~J.}\
  \bibnamefont {Mohr}}, \bibinfo {author} {\bibfnamefont {David~B.}\
  \bibnamefont {Newell}}, \ and\ \bibinfo {author} {\bibfnamefont {Barry~N.}\
  \bibnamefont {Taylor}},\ }\bibfield  {title} {\enquote {\bibinfo {title}
  {Codata recommended values of the fundamental physical constants: 2014*},}\
  }\href {\doibase 10.1103/RevModPhys.88.035009} {\bibfield  {journal}
  {\bibinfo  {journal} {Rev. Mod. Phys.}\ }\textbf {\bibinfo {volume} {88}},\
  \bibinfo {pages} {035009} (\bibinfo {year} {2016})}\BibitemShut {NoStop}%
\bibitem [{\citenamefont {Bliokh}\ \emph {et~al.}(2015)\citenamefont {Bliokh},
  \citenamefont {Rodr{\'\i}guez-Fortu{\~n}o}, \citenamefont {Nori},\ and\
  \citenamefont {Zayats}}]{bliokh2015spin}%
  \BibitemOpen
  \bibfield  {author} {\bibinfo {author} {\bibfnamefont {K~Yu}\ \bibnamefont
  {Bliokh}}, \bibinfo {author} {\bibfnamefont {FJ}~\bibnamefont
  {Rodr{\'\i}guez-Fortu{\~n}o}}, \bibinfo {author} {\bibfnamefont {Franco}\
  \bibnamefont {Nori}}, \ and\ \bibinfo {author} {\bibfnamefont {Anatoly~V}\
  \bibnamefont {Zayats}},\ }\bibfield  {title} {\enquote {\bibinfo {title}
  {Spin--orbit interactions of light},}\ }\href@noop {} {\bibfield  {journal}
  {\bibinfo  {journal} {Nature Photonics}\ }\textbf {\bibinfo {volume} {9}},\
  \bibinfo {pages} {796} (\bibinfo {year} {2015})}\BibitemShut {NoStop}%
\bibitem [{\citenamefont {Stav}\ \emph {et~al.}(2018)\citenamefont {Stav},
  \citenamefont {Faerman}, \citenamefont {Maguid}, \citenamefont {Oren},
  \citenamefont {Kleiner}, \citenamefont {Hasman},\ and\ \citenamefont
  {Segev}}]{Stav1101}%
  \BibitemOpen
  \bibfield  {author} {\bibinfo {author} {\bibfnamefont {Tomer}\ \bibnamefont
  {Stav}}, \bibinfo {author} {\bibfnamefont {Arkady}\ \bibnamefont {Faerman}},
  \bibinfo {author} {\bibfnamefont {Elhanan}\ \bibnamefont {Maguid}}, \bibinfo
  {author} {\bibfnamefont {Dikla}\ \bibnamefont {Oren}}, \bibinfo {author}
  {\bibfnamefont {Vladimir}\ \bibnamefont {Kleiner}}, \bibinfo {author}
  {\bibfnamefont {Erez}\ \bibnamefont {Hasman}}, \ and\ \bibinfo {author}
  {\bibfnamefont {Mordechai}\ \bibnamefont {Segev}},\ }\bibfield  {title}
  {\enquote {\bibinfo {title} {Quantum entanglement of the spin and orbital
  angular momentum of photons using metamaterials},}\ }\href {\doibase
  10.1126/science.aat9042} {\bibfield  {journal} {\bibinfo  {journal}
  {Science}\ }\textbf {\bibinfo {volume} {361}},\ \bibinfo {pages} {1101--1104}
  (\bibinfo {year} {2018})}\BibitemShut {NoStop}%
\bibitem [{\citenamefont {Lodahl}\ \emph {et~al.}(2017)\citenamefont {Lodahl},
  \citenamefont {Mahmoodian}, \citenamefont {Stobbe}, \citenamefont
  {Rauschenbeutel}, \citenamefont {Schneeweiss}, \citenamefont {Volz},
  \citenamefont {Pichler},\ and\ \citenamefont {Zoller}}]{lodahl2017chiral}%
  \BibitemOpen
  \bibfield  {author} {\bibinfo {author} {\bibfnamefont {Peter}\ \bibnamefont
  {Lodahl}}, \bibinfo {author} {\bibfnamefont {Sahand}\ \bibnamefont
  {Mahmoodian}}, \bibinfo {author} {\bibfnamefont {S{\o}ren}\ \bibnamefont
  {Stobbe}}, \bibinfo {author} {\bibfnamefont {Arno}\ \bibnamefont
  {Rauschenbeutel}}, \bibinfo {author} {\bibfnamefont {Philipp}\ \bibnamefont
  {Schneeweiss}}, \bibinfo {author} {\bibfnamefont {J{\"u}rgen}\ \bibnamefont
  {Volz}}, \bibinfo {author} {\bibfnamefont {Hannes}\ \bibnamefont {Pichler}},
  \ and\ \bibinfo {author} {\bibfnamefont {Peter}\ \bibnamefont {Zoller}},\
  }\bibfield  {title} {\enquote {\bibinfo {title} {Chiral quantum optics},}\
  }\href@noop {} {\bibfield  {journal} {\bibinfo  {journal} {Nature}\ }\textbf
  {\bibinfo {volume} {541}},\ \bibinfo {pages} {473} (\bibinfo {year}
  {2017})}\BibitemShut {NoStop}%
\bibitem [{\citenamefont {Ozawa}\ \emph {et~al.}(2019)\citenamefont {Ozawa},
  \citenamefont {Price}, \citenamefont {Amo}, \citenamefont {Goldman},
  \citenamefont {Hafezi}, \citenamefont {Lu}, \citenamefont {Rechtsman},
  \citenamefont {Schuster}, \citenamefont {Simon}, \citenamefont {Zilberberg}
  \emph {et~al.}}]{ozawa2019topological}%
  \BibitemOpen
  \bibfield  {author} {\bibinfo {author} {\bibfnamefont {Tomoki}\ \bibnamefont
  {Ozawa}}, \bibinfo {author} {\bibfnamefont {Hannah~M}\ \bibnamefont {Price}},
  \bibinfo {author} {\bibfnamefont {Alberto}\ \bibnamefont {Amo}}, \bibinfo
  {author} {\bibfnamefont {Nathan}\ \bibnamefont {Goldman}}, \bibinfo {author}
  {\bibfnamefont {Mohammad}\ \bibnamefont {Hafezi}}, \bibinfo {author}
  {\bibfnamefont {Ling}\ \bibnamefont {Lu}}, \bibinfo {author} {\bibfnamefont
  {Mikael~C}\ \bibnamefont {Rechtsman}}, \bibinfo {author} {\bibfnamefont
  {David}\ \bibnamefont {Schuster}}, \bibinfo {author} {\bibfnamefont
  {Jonathan}\ \bibnamefont {Simon}}, \bibinfo {author} {\bibfnamefont {Oded}\
  \bibnamefont {Zilberberg}},  \emph {et~al.},\ }\bibfield  {title} {\enquote
  {\bibinfo {title} {Topological photonics},}\ }\href@noop {} {\bibfield
  {journal} {\bibinfo  {journal} {Reviews of Modern Physics}\ }\textbf
  {\bibinfo {volume} {91}},\ \bibinfo {pages} {015006} (\bibinfo {year}
  {2019})}\BibitemShut {NoStop}%
\end{thebibliography}%

\end{document}